\documentclass[12pt,a4paper]{iopart}

\usepackage{harvard}
\usepackage{epsfig}
\usepackage{color}
\usepackage{graphicx}
\newcommand{\cn}{\citeasnoun}
\newcommand{\np}{\newpage}

\renewcommand{\Eref}[1] {Eq.\,(\ref{#1})}
\newcommand{\ba}{\begin{eqnarray}}
\newcommand{\ea}{\end{eqnarray}}
\newcommand{\be}{\begin{equation}}
\newcommand{\ee}{\end{equation}}
\renewcommand{\br}{\begin{eqnarray*}}
\newcommand{\er}{\end{eqnarray*}}
\newcommand{\la}{\langle}
  \newcommand{\ra}{\rangle}
\usepackage{bm,amssymb}

\newcommand{\Wcm}[2]{
$\rm {#1}\times10^{{#2}}~W/cm^2$}
\newcommand{\bp}{\begin{minipage}}
\newcommand{\ep}{\end{minipage}}
\newcommand{\hs}{\hspace*}
\newcommand{\vs}{\vspace*}

\renewcommand{\ms}{\vs{-5mm}}
\newcommand{\mms}{\vs{-2.5mm}}
\newcommand{\w}{\omega}
\newcommand{\W}{\Omega}
\renewcommand{\l}{\lambda}
\renewcommand{\la}{\langle}
\renewcommand{\ra}{\rangle}
\renewcommand{\r}{\bm r}
\renewcommand{\k}{\bm k}

\newcommand{\nn}{\nonumber}
\renewcommand{\H}{H$_2$~}

\newcommand{\bt}{\begin{tabular}}
\newcommand{\et}{\end{tabular}}

\renewcommand{\Sref}[1] {Sec.~\ref{#1}}

   \renewcommand{\t}{\tau}
\newcommand{\isum}%
{\mathop{\hbox{$\displaystyle\sum\kern-13.2pt\int\kern1.5pt$}}}

\renewcommand{\t}{\tau}

\renewcommand{\b}{\beta}
\newcommand{\q}{\theta}
\newcommand{\Q}{\Theta}
\renewcommand{\l}{\lambda}

\begin{document}

\title[Polarization control of RABBITT in noble gas atoms ]
{Polarization control of RABBITT \\ in noble gas atoms}

\author{ Anatoli S.\ Kheifets$^\dagger$ and Zhongtao Xu}

\address{
	Fundamental and Theoretical Physics, Research School of Physics,\\ 
	The Australian National University, Canberra 2601, Australia
	}
\ead{$^\dagger$a.kheifets@anu.edu.au}
\vspace{10pt}
\ms
\begin{abstract}

The mutual angle formed by the non-collinear polarization axes of two
laser pulses is used to control two-photon XUV+IR ionization of noble
gas atoms in the process of reconstruction of attosecond bursts by
beating of two-photon transitions (RABBITT).  The magnitude and the
phase of this beating can be controlled very efficiently by the mutual
polarization angle.  The mechanism of this control can be understood
within the lowest order perturbation theory and the soft photon
approximation. We offer a very sensitive test on the polarization
control of the angular dependent RABBITT process which validates
our numerical simulations. We apply this test to the recent
theoretical and experimental results of polarization controlled
RABBITT on hydrogen and helium by Boll {\em et al.}, Phys. Rev. A 107,
043113 (2023) and  heavier noble gases by Jiang {\em~et~al.},
Nature Comms. 13, 5072 (2022).

\end{abstract}

\date{\today}

\section{Introduction} 

Two-color two-photon extreme ultraviolet and infrared (XUV+IR)
photoionization has been applied recently for studying ultrafast
electron dynmaics on the attosecond time scale.  Reconstruction of
attosecond bursts by beating of two-photon transitions (RABBITT)
\cite{PaulScience2001,Mairesse1540} is one practical realization of
this technique. In RABBITT, two collinearly polarized XUV and IR laser
pulses with a variable delay are used to ionize the target atom and to
steer emitted photoelectrons. The two-photon ionization yield
oscillates with twice the IR photon frequency as the XUV/IR pulse delay
varies. The phase of this oscillation encodes the timing of the XUV
ionization \cite{PhysRevA.54.721,Dahlstrom201353}. Both the phase and
magnitude of the RABBITT oscillation depend sensitively on the
photoelectron escape angle relative to the common polarization axis of
the XUV and IR pulses
\cite{PhysRevA.94.063409,PhysRevA.96.013408,PhysRevA.97.063404}.

An additional control of two-color photoionization can be gained by
relaxing the IR polarization direction and allowing its rotation
relative to the XUV polarzation axis
\cite{OKeeffe2004,Meyer2008,Meyer2010,Leitner2015,Boll2020}. Recently,
such a polarization control was implemented in RABBITT.
\cn{Jiang2022} demonstrated the so-called ``atomic partial wave
meter'' where non-collinear partial waves with magnetic projections
$M\ne0$ increase their presence gradually as the mutual polarization
axes angle grows.  \cn{Boll2023} demonstrated an appearance of an
additional set of angular nodes of the RABBITT amplitude in
$s$-electron targets (H and He).

In the present work, we expand these initial investigations of
polarization controlled RABBITT. In our theoretical modeling we use a
numerical solution of the time-dependent Schr\"odinger equation (TDSE)
to generate the angular-dependent phase and magnitude of the RABBITT
oscillations in helium, neon and argon. In addition, we employ the
lowest order perturbation theory (LOPT)
\cite{PhysRevA.54.721,Dahlstrom201353} and the soft photon
approximation (SPA) \cite{Maquet2007} to provide qualitative
interpretation of our numerical results.  While we confirm the angular
node structure of the RABBITT amplitude in $s$-electron targets (H and
He) as reported by \cn{Boll2023}, the additional nodes are missing in
heavier noble gases (Ne and Ar). This is explained within the SPA in
which the RABBITT amplitude possesses the angular symmetry of XUV
ionization $1+\beta P_2(\cos\theta)$ modulated by the IR angular
factor $\cos^2(\q-\Q)$.  Here $\q$ is the photoelectron escape angle
and $\Q$ is the IR polarization axis angle. Both angles are counted
relative to the XUV polarization direction which is taken as the
quantization axis.  In $s$-electron targets (H and He) the angular
anisotropy parameter $\beta=2$ and the XUV angular factor reduces to
$\cos^2\theta$ which possesses an angular node. In $p$ valence shells
of heavier noble gases (Ne and Ar), $\b<2$ and the XUV angular factor
remains finite. Owing to the XUV+IR angular structure, the angular
dependent RABBITT phase in H and He remains symmetric relative to the
angle $\q-\Q/2=90^\circ$ whereas it shows the angular symmetry
relative to the angle $\q-\Q=90^\circ$ in Ne and Ar at small
photoelectron energies where $|\b|\lesssim1$.

We confirm the gradual increase of partial waves with $M\ne0$ towards
the orthogonal field configuration $\Q\to\pi/2$ when the IR angular
factor reduces to $\sin\q\propto Y_{11}(\q)$. This increase is
particularly graphical in $s$-electron targets.  In all the targets,
the photoelectron momentum distribution (PMD) displays a systematic angular
rotation with the polarization control angle $\Q$.

The rest of the paper is organized in the following way. In
\Sref{Theory} we give a brief outline of our analytic tools
(\Sref{Analytic}) and  computational methods
(\Sref{Numeric}).  \Sref{Results}
contains our main numerical results presented consequently for helium,
neon and argon. We conclude in \Sref{Conclusion} by outlining further
extensions of the present study.

\mms
\section{Theory}
\label{Theory}

\mms
\subsection{Analytic formulation}
\label{Analytic}

In a typical RABBITT measurement, an ionizing XUV attosecond pulse
train (APT) is superimposed on an attenuated and variably delayed
replica of the driving IR pulse.  The XUV photon $\W=(2q\pm1)\w$ is
absorbed from the initial bound state and then is augmented by an IR
photon absorption $+\w$ or emission $-\w$ leading to formation of the
even order sideband (SB) in the photoelectron spectrum.  The center of
the IR pulse is shifted relative to the APT by a variable delay $\tau$
such that the magnitude of a SB peak oscillates as
\be
S_{2q}(\tau) =
A+B\cos[2\omega\tau-C]
\ .
\label{oscillation}
\ee
The RABBITT parameters $A$, $B$ and $C$ entering
Eq.~(\ref{oscillation}) can be expressed as
\ba
\nn
A&=&\sum_{m_i}|{\cal M}^{(-)}_{m_i}(\k)|^2+|{\cal M}^{(+)*}_{m_i}(\k)|^2
\  , \ 
B=2{\rm Re}\sum_{m_i} \left[{\cal M}^{(-)}_{m_i}(\k)
{\cal M}^{(+)*}_{m_i}(\k)\right]
\\
C&=& \arg\sum_{m_i}\left[
{\cal M}^{(\rm -)}_{m_i}(\k)
{\cal M}^{*(\rm +)}_{m_i}(\k)
\right]
\equiv 2\w\tau_a
\ .
\label{abc}
\ea
Here ${\cal M}^{(\pm)}_{m_i}(\k)$ are complex and angle-dependent
amplitudes of two-photon ionization produced by adding $(+)$ or
subtracting $(-)$ an IR photon, respectively. An incoherent summation
over the angular momentum projection of the initial state $m_i$ is
explicit in \Eref{abc}.  The atomic time delay $\tau_a$ quantifies the
timing of the XUV ionization process.

By adopting the soft photon approximation (SPA)
\cite{Maquet2007} we can write
\ba
\label{soft}
A,B &\propto&
|J_1({\bm \alpha}_0\cdot\k)|^2
|\la f|z|i \ra|^2
\propto
\left[1+\beta P_2(\cos\theta)\right]
(\hat{\bm\alpha}\cdot\hat{\k})^2  \ .
\ea
Here we make a linear approximation to the Bessel function as the
parameter $ \alpha_0= F_0/\omega^2\ll1$ in a weak IR field with a
small magnitude $F_0$.  The angular anisotropy $\b$ parameter defines
the photoelectron angular distribution in single-photon XUV
ionization.  Derivation of \Eref{soft} can be found in the Appendix of
\cn{PhysRevA.97.063404}. Similar equations are used e.g. by
\cn{Boll2017}.

The angular dependence of the amplitudes ${\cal M}_{\k}^{\pm}$ can be
deduced from the LOPT expression \cite{Dahlstrom201353}:
\ba
\label{lopt}
\nn
\hs{-2.5cm}
{\cal M}^{\pm}_{m_i}(\k) &\hs{-1.cm}\propto&\hs{-0.5cm} 
\sum_{\l=l_i\pm1}
\sum_{L=\l\pm1}
\sum_{|M|\leq L ; \mu|\leq\l}
(-i)^L e^{i\eta_L} Y_{LM}(\hat{k})
\isum \ d^3\kappa \
{\la R_{kL}|r|R_{\kappa\l}\ra \la R_{\kappa\l}|r|R_{l_in_i}\ra\over
  E_i+\W^\pm-\kappa^2/2-i\gamma}
\\
&&\hs{2cm}\times
\la Y_{LM}|\hat{\bm\alpha}\cdot\hat{\k}|Y_{\l\mu}\ra 
\la Y_{\l\mu}|\cos\theta|Y_{l_im_i}\ra 
\ea
In the above expression, $\la n_il_i|, \la \kappa\l|$ and $\la kL|$ are
the initial, intermediate and final electron states defined by their
linear and angular momenta, the latter are bound by the triangular
angular momentum coupling rule. The XUV photon energy is
$\W^\pm=(2q\pm1)\w$ and $i\gamma$ denotes the pole bypass in the
complex energy plane. In the collinear case, $M=0$ and only the
axially symmetry spherical harmonics $Y_{L0}$ make their contribution.

To evaluate the angular integrals in \Eref{lopt} we transform the
trigonometric functions into the matching spherical
harmonics \footnote{The corresponding evaluations are documented in
  the Mathematica notebook that can be found in the Supplementary
  Material.}. The
product integral of three spherical harmonics is known analytically
\cite{V88}.  In the simplest case of an $s$-electron target, $m_i=0$, $\l=1$
and $L=\{0,2\}$ which takes us to Eq.~(3) of
\cn{Jiang2022}\footnote{The original equation of \cn{Jiang2022}
  differs by an extra magnitude factor of $1/\sqrt3$ and an
  alternative selection of the polar angle $\phi\to\phi+\pi/2$}:
\be
\label{angular}
\hs{-2cm}
{\cal M}_{\k}^{\pm} \propto
\cos\Q
\left[\frac{1}{\sqrt 3} Y_{00}(\hat k) \ T^\pm_0
+\frac{2}{\sqrt{15}}Y_{20}(\hat k) \ T^\pm_2\right]
+ 
\sin\Q\frac{1}{\sqrt{10}}
[Y_{21}(\hat k)-Y_{2-1}(\hat k)] \ T^\pm_2
\ee
Here we reintroduced the radial factors $T^\pm_L$ following
notations of \cn{Boll2023}.  The angular nodes of ${\cal
  M}_{\k}^{\pm}$ correspond to the photoelectron emission angles
$\q^\pm$ which satisfy Eqs.~(11) and (12) of \cn{Boll2023}:
\be
\label{node2}
\hs{-2cm}
\cos(2\q^\pm-\Q) = -\left[
\frac13 +\frac23 {T^{\pm}_0 \over T^{\pm}_2} 
\right]\cos\Q
\ \ , \ \ 
\cos2\q^\pm = -\left[
\frac13 +\frac23 {T^{\pm}_0 \over T^{\pm}_2} 
\right]
\ \ {\rm for} \  \Q=0
\ee
In the collinear case $\Q=0$, these nodes are controlled by the ratio
$T^{\pm}_0 / T^{\pm}_2$. In the non-collinear case, additional nodes
appear which depend on the polarization control angle $\Q$. 

For a $p$-electron target, an angular node in 
${\cal   M}^{\pm}_{m_i}(\k)$ 
at a given $m_i$ does not match nodes at other $m_i$ values. For
instance, in the simplest collinear $\Q=0$ case, the node equations
read
\be
\cos(2\q^\pm) =
\left\{ 
\begin{array}{ccc}
 \frac15 \left[
1 - \frac83 \
{T^\pm_1\over T^\pm_3} \right] 
\to - \frac13
\ \ {\rm as } \ \
T^\pm_1\to  T^\pm_3
& {\rm for} \ \ m_i=0\\
 - \frac35\left[
1 + \frac23
{T^\pm_1\over T^\pm_3} \right]
\to -1
\ \ {\rm as } \ \
T^\pm_1 \to  T^\pm_3
& {\rm for} \ \ m_i=1
\end{array}
\right.
\ee
In the $m_i=0$ case, the nodes approach asymptotically the magic angle
of $54.7^\circ$ whereas for $m_i=1$ these nodes tend asymptotically to
$90^\circ$. Because of this mismatch, the magnitude $B,C$ parameters
remain nodeless.

\subsection{Numerical methods}
\label{Numeric}

We follow closely the previous works by \cn{PhysRevA.97.063404} and
\cn{PhysRevA.103.L011101}. In brief, we solve numerically the
time-dependent Schr\"odinger equation (TDSE) in a single-active
electron approximation:
\footnote[1]{ Here and throughout, we use the atomic
  units (a.u.) by setting $e=m=\hbar=1$.}
\begin{equation}
\label{TDSE}
i {\partial \Psi(\r) / \partial t}=
\left[\hat H_{\rm atom} + \hat H_{\rm int}(t)\right]
\Psi(\r) \ .
\end{equation}
Here the radial part of the atomic Hamiltonian
\be
\label{Hat}
\hat H_{\rm atom}(r) = 
-\frac12{d^2\over dr^2} +{l(l+1)\over 2r^2} + V(r)
\ee
contains an effective one-electron potential $V(r)$ obtained by
localization of the non-local Hartree-Fock potential
\cite{0022-3700-11-24-007}.  The Hamiltonian $\hat H_{\rm int}(t)$
describes interaction with the external field and is written in the
velocity gauge
\be
\label{gauge}
\hat H_{\rm int}(t) =
 {\bm A}(t)\cdot \hat{\bm p} \ \ , \ \ 
{\bm A(t)}=-\int_{0}^{t}{\bm E(t')}\ d t' \ .
\ee
This external field is comprised of combination of XUV and IR fields.
The XUV field is represented by  an APT modeled by the 
vector potential
\ba
\label{vectorGauss}
A_x(t) &=& \sum_{n=-5}^5 (-1)^n A_n \exp\left(
-2\ln2
{(t-nT/2)^2\over \tau_x^2}
\right)\nonumber\\
&&\times
\cos\Big[\omega_x(t-nT/2)\Big]  
\ea
with the magnitude of an $n$th pulselet
$$
A_n = A_0
\exp\left(-2\ln2
{(nT/2)^2\over \tau_T^2}\right) \ .
$$
Here $A_0$ is the vector potential peak value
%
%
and $T=2\pi/\omega$ is the period of the IR field.
The vector potential of the IR pulse is defined as
\be
\label{vectorSin2}
A(t) = A_0 \cos^2
\left(
{\pi (t-\tau)\over 2\tau_{\rm IR}}
\right)
\cos[\omega(t-\tau)] \ .
\ee
We select the fundamental IR frequency $\w=1.55$~eV corresponding to the
wavelength $\l=800$~nm.  The XUV central frequency is set to
$\omega_x=15\w$ and the APT parameters are chosen as $\tau_T=5$~fs and
$\tau_x=30$~as. The IR pulse length is defined by $\tau_{\rm IR}=15$~fs.
The intensity of the IR and XUV pulses is chosen in the range of
\Wcm{1}{10}. Such an intensity is sufficiently low to stay within the
LOPT boundaries.

The  TDSE \eref{TDSE} is solved numerically  by  the
spherical-coordinate implicit derivatives (SCID) computer code
\cite{PATCHKOVSKII2016153}. The PMD is obtained using the
time-dependent surface flux (t-SURF) method
\cite{TaoNJP2012,0953-4075-49-24-245001}. The PMD can be expressed as
the projection of the time-dependent wave function after the end of
propagation on the basis of scattering states
\be
\label{PMD}
P_{nl}(\k) = \sum_m
\Big|\la \varphi_{\k}(r)|\Psi_{nlm}(\r,t\to\infty)|\ra\Big|^2
\ .
\ee
Here the indexes $n,l,m$ denote the initial atomic bound state. The
photoelectron momentum $\k$ is defined in the Cartesian frame in which
the $\hat z$ axis is aligned with the polarization vector of the XUV
pulses and the $(x,z)$ plane contains both the polarization vectors of
the XUV and IR pulses.  Projection of the PMD on this plane serves to
determine the angular profile of a given spectral feature:
\be
\label{angular}
\hs{-1.5cm}
S_{N}(\theta) =    P_{nl}(k_x,k_y=0,k_z)
\ \ , \ 
\theta = \tan^{-1}(k_z/k_x)
\ \ , \ 
(k_x^2+k_z^2)/2 = N\w-I_p
\ .
\ee
Here $N=2q$ for a sideband and $N=2q\pm1$ for a primary harmonic peak.
In \Eref{angular} the bound state indices $nl$ are assumed on
$S_N(\q)$ but dropped for brevity and $I_p$ is the ionization
potential of the target orbital. The $k$-grid is sufficiently dense to
specify the angle $\q$ with a $2^\circ$ increment.

\begin{figure}
\epsfxsize=7cm
\epsffile{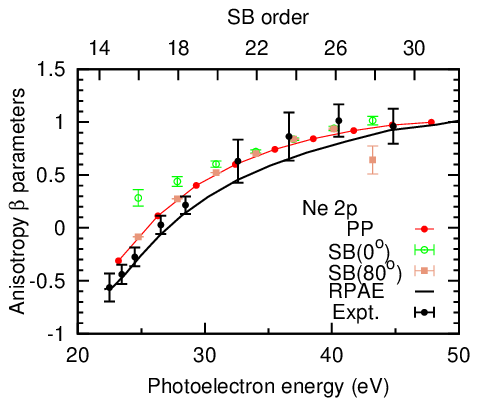}
\epsfxsize=7cm
\epsffile{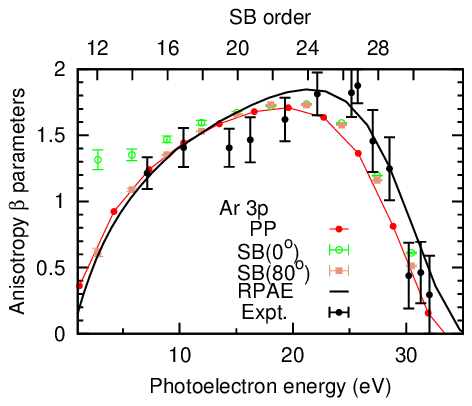}

\caption{Angular anisotropy $\b$ parameter in Ne (left) and Ar (right)
  as extracted from the angular fitting to the primary harmonic peaks
  (PP), sidebands (SB) and the RPAE calculation. The SB data
  correspond to the polarization control angle of $\Q=0$ and
  $\Q=80^\circ$. The experimental data are from \cn{0022-3700-9-5-004}
  for Ne and \cn{0022-3700-7-17-003} for Ar.
\label{Fig1}}
\ms
\end{figure}

The IR pulse is systematically shifted relative to the APT by a
variable delay $\tau$ in six increments by 10~a.u. This induces a
time oscillation of the magnitude of each sideband $S_{2q}$ which is
fitted with \Eref{oscillation} to obtain the angular dependent
magnitude ($A,B$) and phase ($C$) parameters. The magnitude parameter $B$
is then fitted with \Eref{soft} to deduce the angular anisotropy $\b$
parameter. The same parameter can also be extracted from the angular
variation of the prime harmonic peaks
\be
\label{conventional}
S_{2q\pm1}(\theta) \propto
1+\b P_2(\cos\theta)
\vs{1mm}
\ee
In addition, the $\b$ parameter is calculated in the time-independent
random phase approximation with exchange (RPAE) \cite{A90}. Thus
obtained $\b$ parameters in Ne and Ar are displayed in the left and
right panels of \Fref{Fig1}, respectively. The SB($0^\circ$) and
SB($80^\circ$) $\b$ parameters correspond to $\Q=0$ and
$80^\circ$. The latter set is in better agreement with other $\b$
values for both target atoms. 
This indicates that the validity of the
SPA improves as the polarization control angle grows. The same
conclusion was reached by \cn{Boll2023}. 

\section{TDSE results}
\label{Results}

\subsection{Helium}

\begin{figure}[ht]
\bp{20cm}
\hs{-2cm}
\epsfxsize=7cm
\epsffile{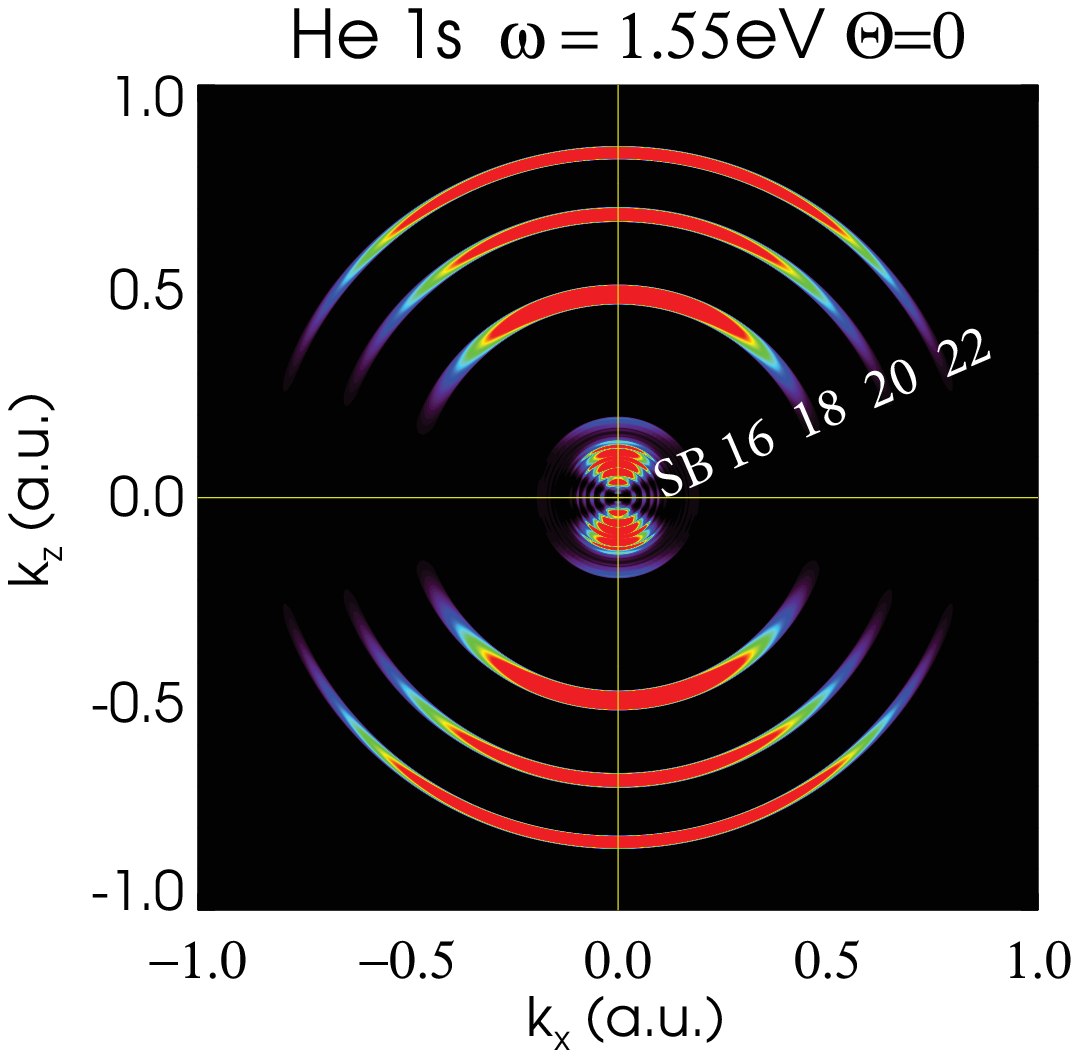}
\hs{-3.3cm}
\epsfxsize=7cm
\epsffile{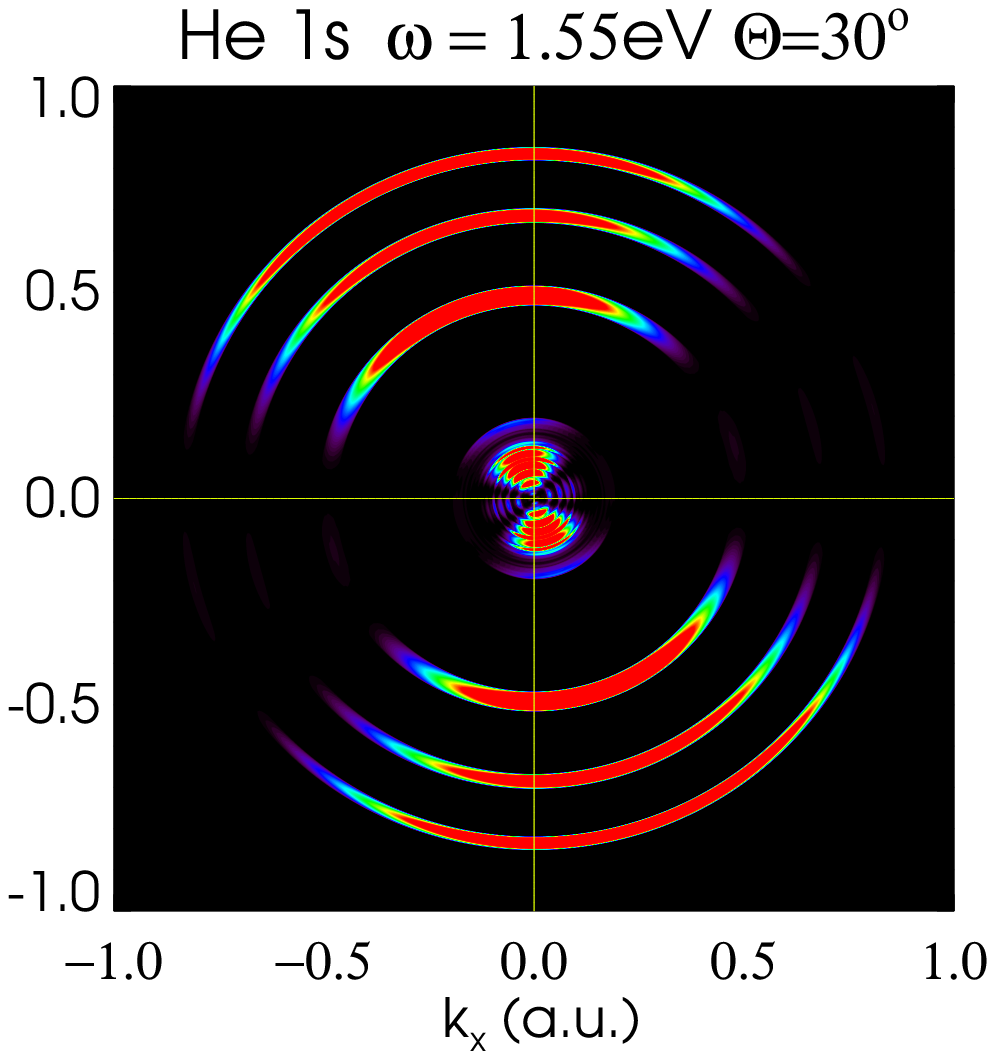}
\hs{-3.3cm}
\epsfxsize=7cm
\epsffile{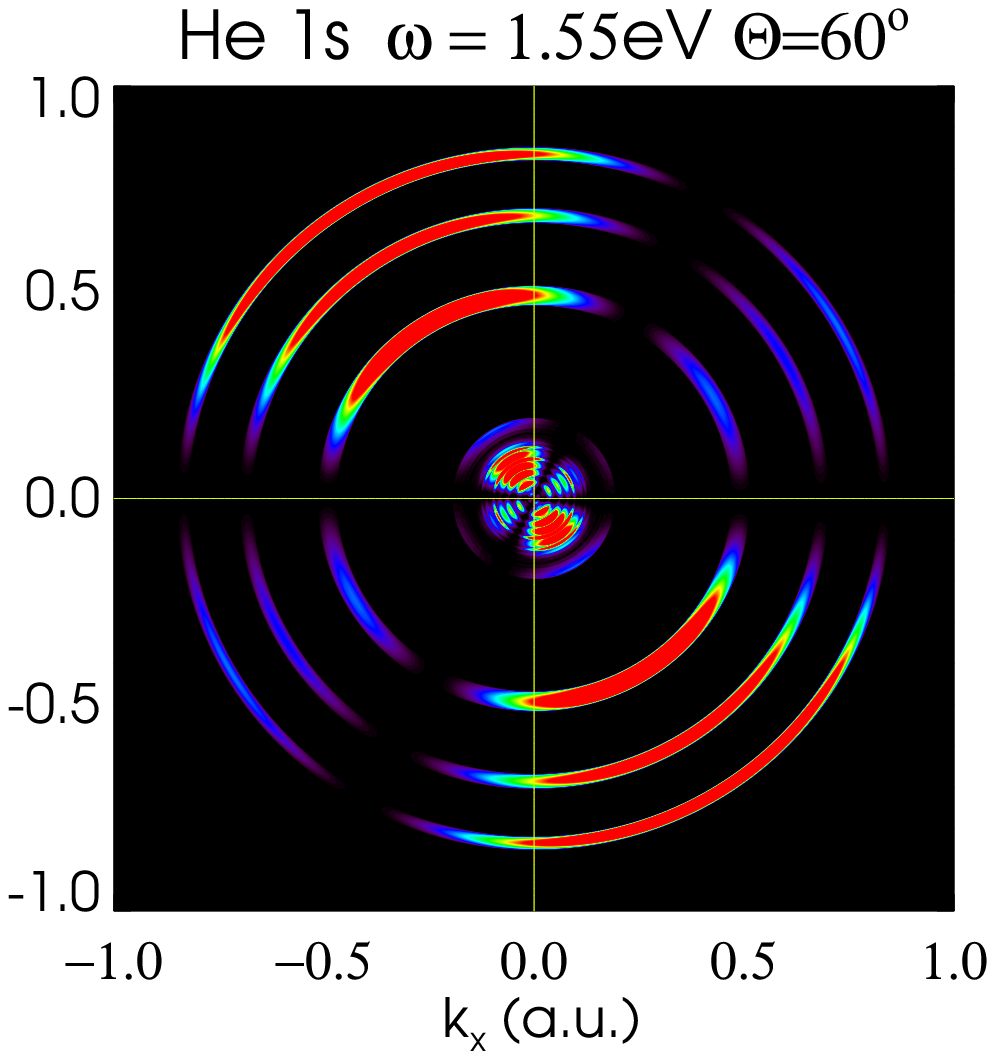}
\hs{-3.3cm}
\epsfxsize=7cm
\epsffile{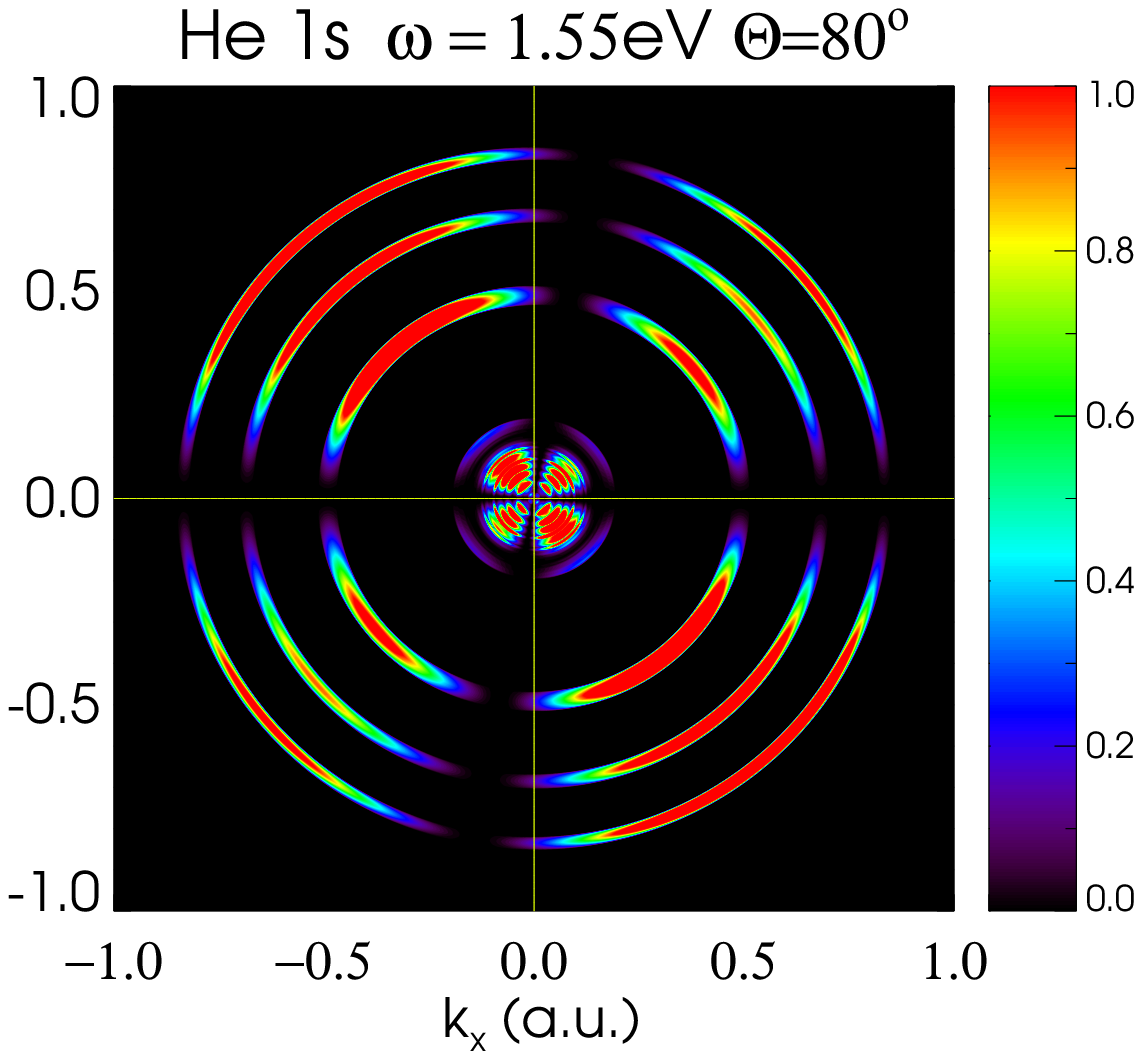}

\hs{-0.3cm}
\epsfxsize=16.5cm
\epsffile{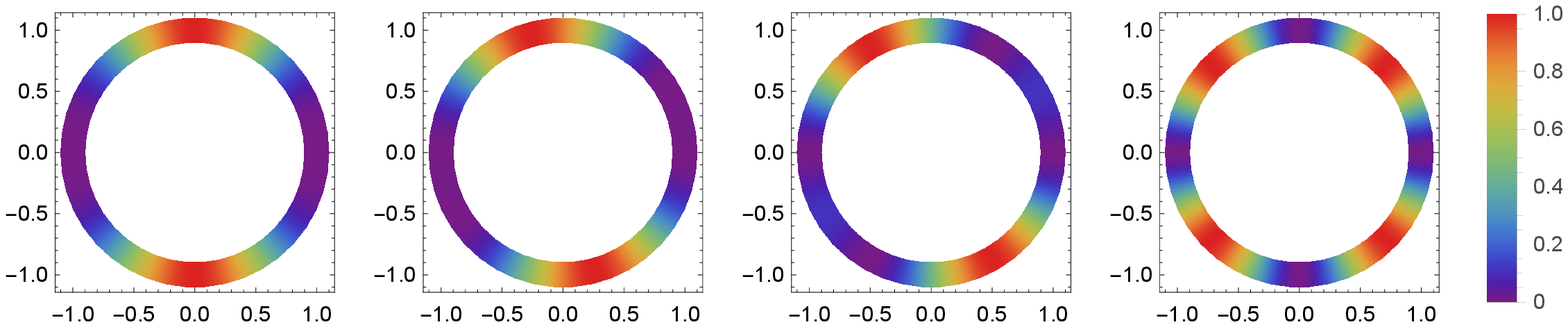}
\ep

\caption{Top row: PMD of helium projected on the joint XUV/IR
  polarization plane at various angles $\Q$. The XUV/IR time delay
  $\t=0$ in all cases. The SB orders are as
  marked. Bottom row: graphical visualization of the angular factor
  $\cos^2(\q-\Q)\cos^2\q$.
\label{Fig2}}
\end{figure}

The PMD of He projected on the joint polarization plane is exhibited
in the top row of panels in \Fref{Fig2}.  Each panel corresponds to a
fixed polarization control angle $\Q$.  The sidebands are highlighted in the
figure by applying a band pass filter
\be
\label{filter}
\hs{-2cm}
\overline P(k_x,k_z) = \sum_{2q}\int k^2\,dk \  P(k_x,k_y=0,k_z)
\ , \ \ 
E_{2q}-\Gamma/2<k_x^2+k_z^2<E_{2q}+\Gamma/2
\ ,
\ee
The filtered PMD \eref{filter} is zero outside the spectral width of
the sideband $\Gamma$. Such a filtering masks the normally dominant
prime harmonic peaks which are not sensitive to the polarization
control.  In the case of an $s$-electron targets (H and He), $\b=2$
and the angular factor of \Eref{soft} is reduced to
$\propto\cos^2\q\cos^2(\q-\Q)$. This factor is visualized on the
bottom row of panels in \Fref{Fig2}.  The sidebands at $\Q=0$ have the
$\cos^4\theta\propto|Y_{20}|^2$ angular pattern while at
$\Q\to90^\circ$ the sidebands acquire the
$\cos^2\theta\sin^2\theta\propto|Y_{21}|^2$ symmetry. This is in line
with the ``atomic partial wave meter'' effect demonstrated by
\cn{Jiang2022}.

We note that the primary harmonic peak PP15 submerges below the
threshold in He. Correspondingly, the SB16 is formed by the
under-threshold uRABBITT process. This process has been studied
extensively in He
\cite{SwobodaPRL2010,Drescher2022,Neoricic2022,Autuori2022} and
heavier noble gases - Ne
\cite{Villeneuve2017,PhysRevA.103.L011101,Kheifets2021Atoms} and Ar
\cite{Kheifets2023}. While we do not observe a noticeable deviation of
the angular symmetry of SB16 from other sidebands in He, this symmetry
may differ in heavier noble gases.  

\np

\begin{figure}[ht]
\bp{20cm}
\hs{-1cm}
\epsfxsize=6cm
\epsffile{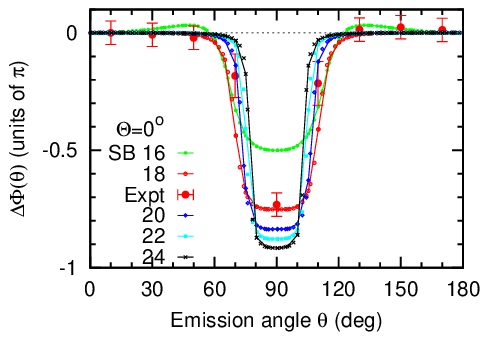}
\hs{-1cm}
\epsfxsize=6cm
\epsffile{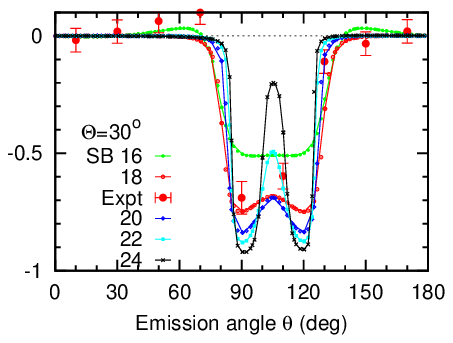}
\hs{-1cm}
\epsfxsize=6cm
\epsffile{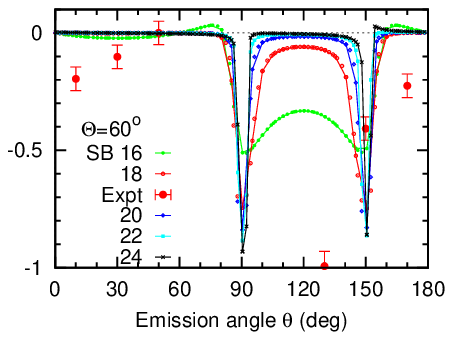}

\hs{-1cm}
\epsfxsize=6cm
\epsffile{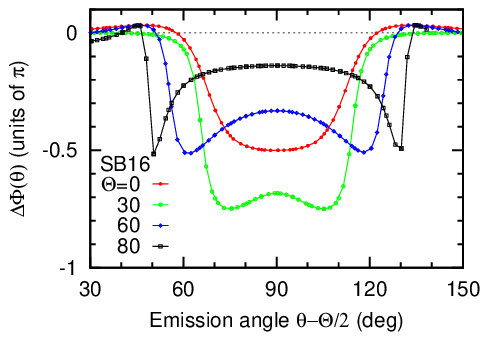}
\hs{-1cm}
\epsfxsize=6cm
\epsffile{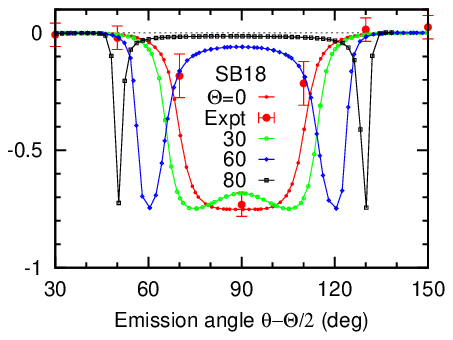}
\hs{-1cm}
\epsfxsize=6cm
\epsffile{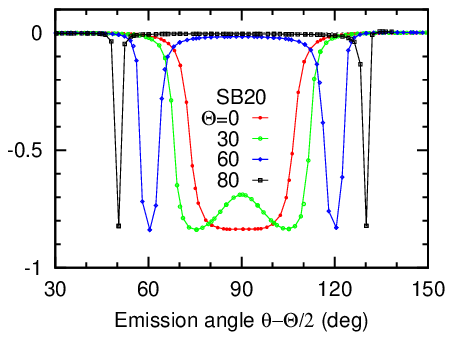}

\hs{-1cm}
\epsfxsize=6cm
\epsffile{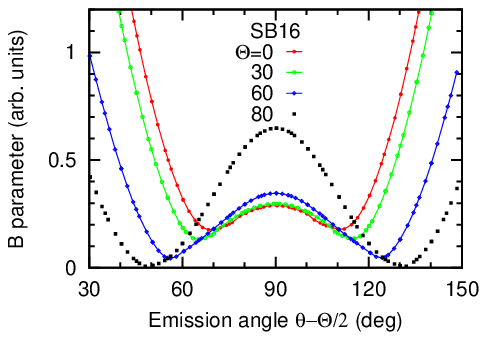}
\hs{-1cm}
\epsfxsize=6cm
\epsffile{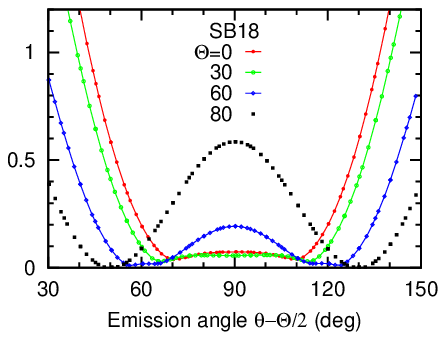}
\hs{-1cm}
\epsfxsize=6cm
\epsffile{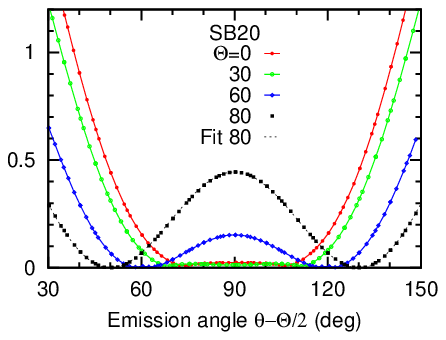}
\ep

\caption{Angular dependence of RABBITT parameters in helium. Top row:
  RABBITT phase ($C$-parameter) counted relative to the zero emission
  angle $\Delta\Phi(\q)=\Phi(\q)-\Phi(0)$ is shown for various SB
  orders at a fixed polarization control angle $\Q$. The experimental
  data of \cn{Jiang2022} for SB18 at the polarization control angles
  $\Q=0$, $20^\circ$ and $54.7^\circ$ are shown with error
  bars. Middle row: $\Delta\Phi$ is shown for several fixed SB orders
  while $\Q$ angle varies. The horizontal emission angle scale is
  shifted as $\q-\Q/2$. Bottom row: $B$ parameter is shown for several
  fixed SB orders while $\Q$ angle varies. The dotted line shows the
  analytic fit with $\cos^2\q\cos^2(\q-\Q)$ for $\Q=80^\circ$.
\label{Fig3}}
\end{figure}

Another set of the polarization control data is displayed in \Fref{Fig3}
where we show the angular dependent RABBITT phase ($C$-parameter)
and magnitude ($B$-parameter). The relative phase is 
counted relative to the XUV polarization direction
$\Delta\Phi(\q)=\Phi(\q)-\Phi(0)$. In each panel of the top row, we
display $\Delta\Phi$ for a range of SB orders at a fixed $\Q$
value. Our results reproduce very closely the analogous set of data
for hydrogen exhibited in Figs.~2 and 3 of \cn{Boll2023}.
Although hydrogen is not a noble gas atom, the RABBITT process in H is
very similar to that in He with a small adjustment of the
photoelectron energy to account for different ionization potentials
\cite{PhysRevA.94.063409,Boll2020,Boll2023}.

The characteristic feature of the angular dependent RABBITT phase is
its jump by about one unit of $\pi$ which is accompanied by a drop of
the RABBITT magnitude. An angular node of the magnitude would have
induced an exact $\pi$ phase jump. In the collinear case $\Q=0$, the
normally flat phase starts deviating from its polarization direction
value near the ``magic'' angle of $54.7^\circ$. At this angle, the
normally dominant $d$-wave passes through a kinematic node and the
normally weak $s$-wave takes over.  The phase jump becomes steeper as
the SB order grows. This is explained by a gradual convergence of the
$T_0$ and $T_2$ factors in \Eref{node2}. The non-collinear case
$\Q\ne0$ is markedly different as the phase tends back to its $\q=0$
value after the jump. This tendency is particularly clear at large
angles $\Q$ when the second upwards phase jump occurs almost
immediately after the first downwards jump. The latter behavior can be
explained by the gradual increase of the $M=\pm1$ components in the
final continuum which is not supported by the $s$-wave. Hence the
$d$-wave remains dominant everywhere except for an immediate vicinity
of its kinematic node. The experimental phases of \cn{Jiang2022} for
SB18 show a good agreement with our calculation in the collinear case
$\Q=0$. However this agreement becomes poorer as the polarization
control angle $\Q$ increases. The sparsity of the experimental data
points does not allow to reveal the secondary phase jumps at large
$\Q$ angles.

In the middle row of panels, we replot $\Delta\Phi$ by adjusting the
horizontal scale and shifting the photoelectron emission angle by
$\Q/2$. With this shift, all the angular distributions become
perfectly centered with respect to the emission angle
$\q-\Q/2=90^\circ$. Such a rigid alignment can be explained very
simply by the RABBITT angular factor
$
\cos(\theta-\Q)\cos\theta
\propto \cos(2\theta-\Q)\propto\cos(\theta-\Q/2)
\ .
$

We note that the SB16 phase behavior as a function of the polarization
control angle $\Q$ is noticeably different from that of higher order
SB18 and SB20. The phase jump in SB16 is markedly smaller and the
phase never returns to its $\q=0$ value. This is another manifestation
of the uRABBITT effect. 
We also note that the experimental data of \cn{Jiang2022} for SB18
tend to cluster around the center at $\q-\Q/2=90^\circ$ while the
accompanying calculations of the same authors as well as that of
\cn{Boll2023} follow  this pattern more clearly.

The bottom row of panels in \Fref{Fig3} displays the angular dependent
$B$ paramters on the adjusted angular scale. On this scale, the $B$
parameters are perfectly centered relative to the angle
$\q-\Q/2=90^\circ$. In the SB16 formed by the uRABBITT process, the
$B$ parameter falls sharply but does not reach zero at $\Q=0$ and
$30^\circ$. Accordingly, the phase displays a rather smooth angular
variation by less than one unit of $\pi$. At larger angles $\Q=60$ and
$80^\circ$, an additional node appears and the phase displays a rather
sharp oscillation. In the two conventional SB18 and SB20, the $B$
parameters drop much closer to zero at $\Q=0$ and $30^\circ$ and the
phase $C$ parameter displays a steeper variation closer to one unit of
$\pi$. The angular nodes are fully formed at $\Q=60$ and $80^\circ$
where they follow very closely the angular pattern
$\cos^2\q\cos^2(\q-\Q)$.  Accordingly, the phase oscillation becomes
very sharp.

\begin{figure}[h]
\bp{20cm}
\hs{-2cm}
\epsfxsize=7cm
\epsffile{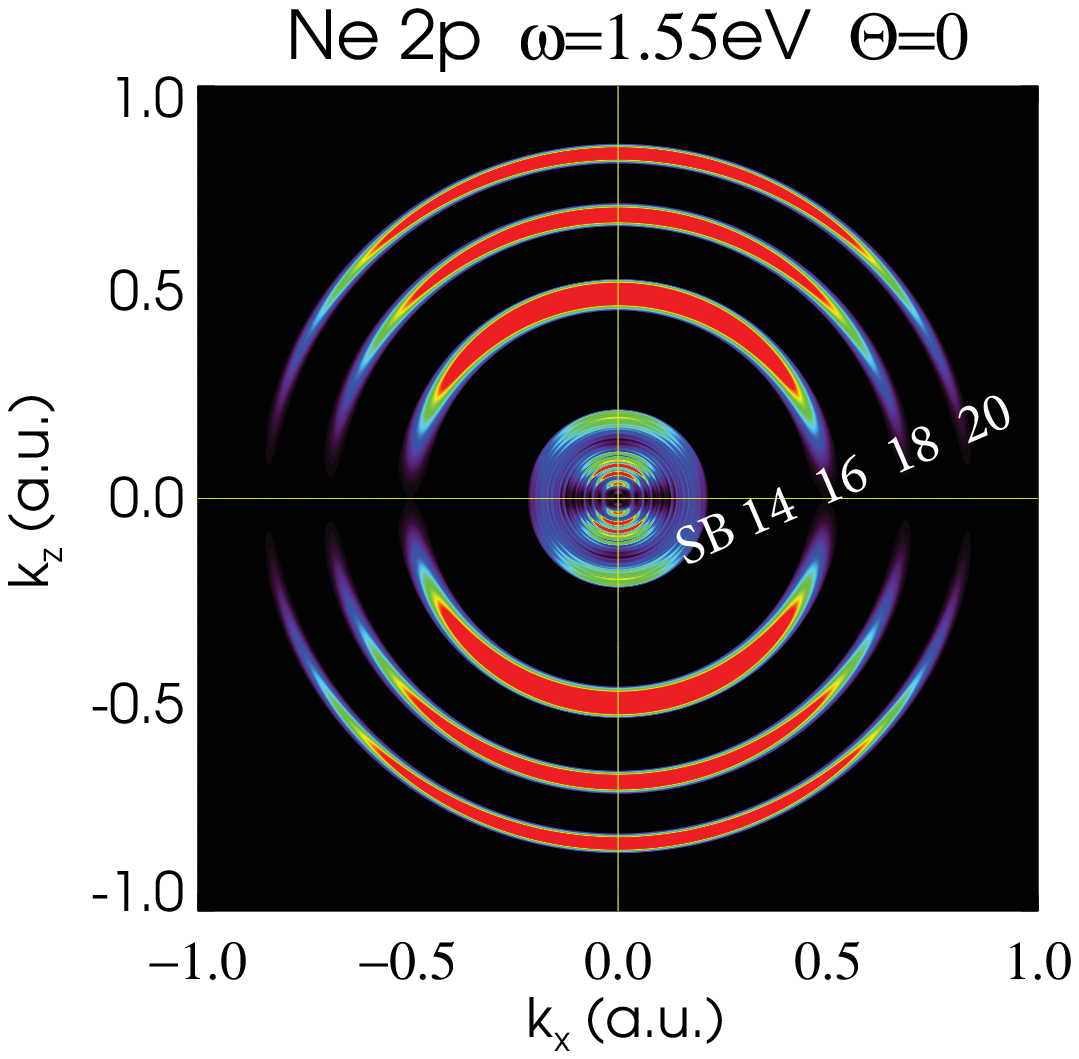}
\hs{-3.3cm}
\epsfxsize=7cm
\epsffile{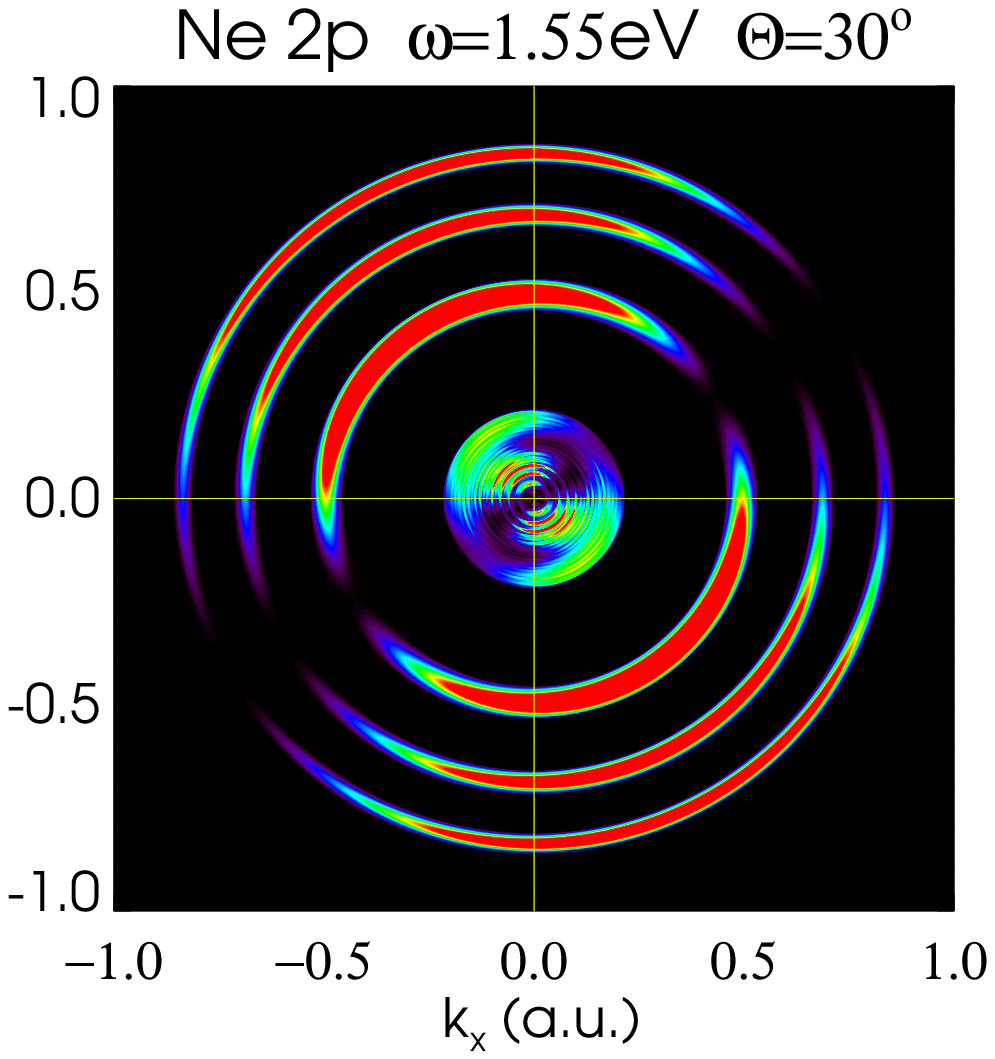}
\hs{-3.3cm}
\epsfxsize=7cm
\epsffile{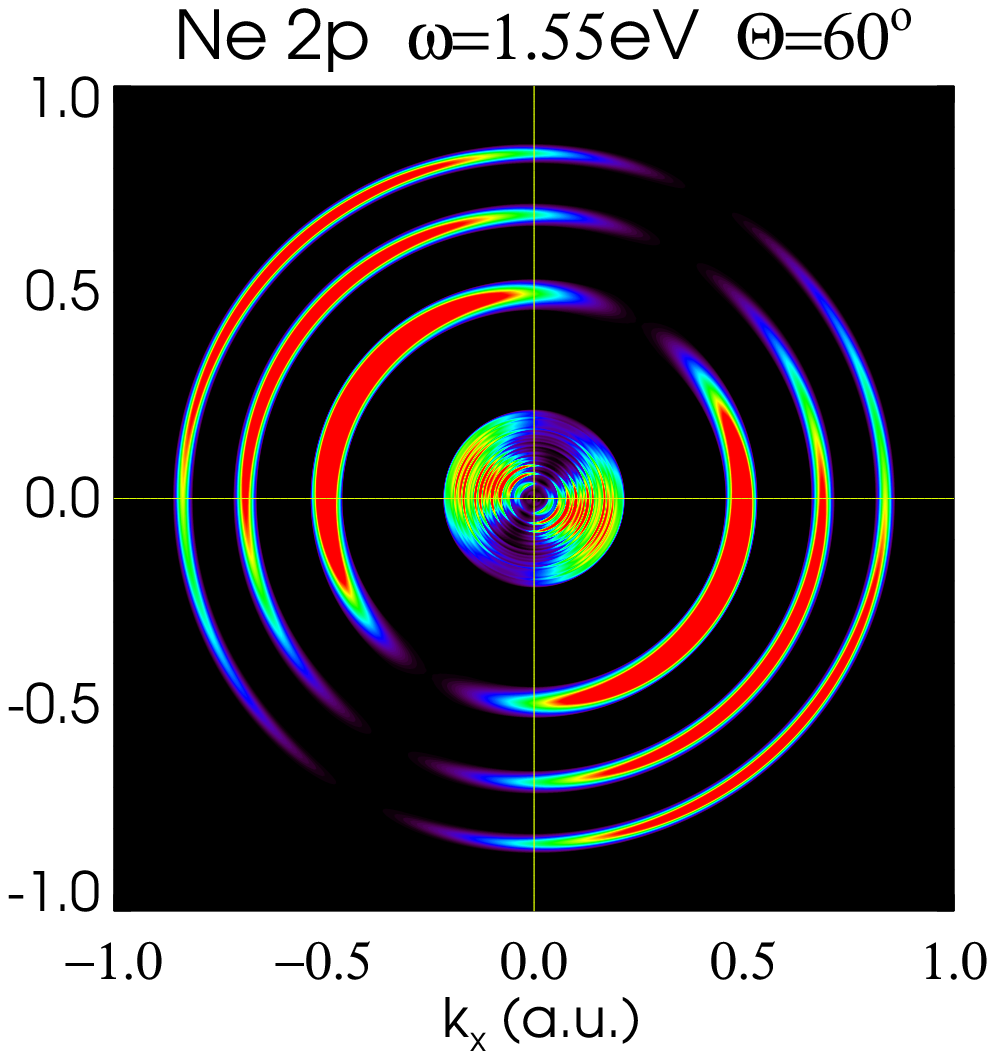}
\hs{-3.3cm}
\epsfxsize=7cm
\epsffile{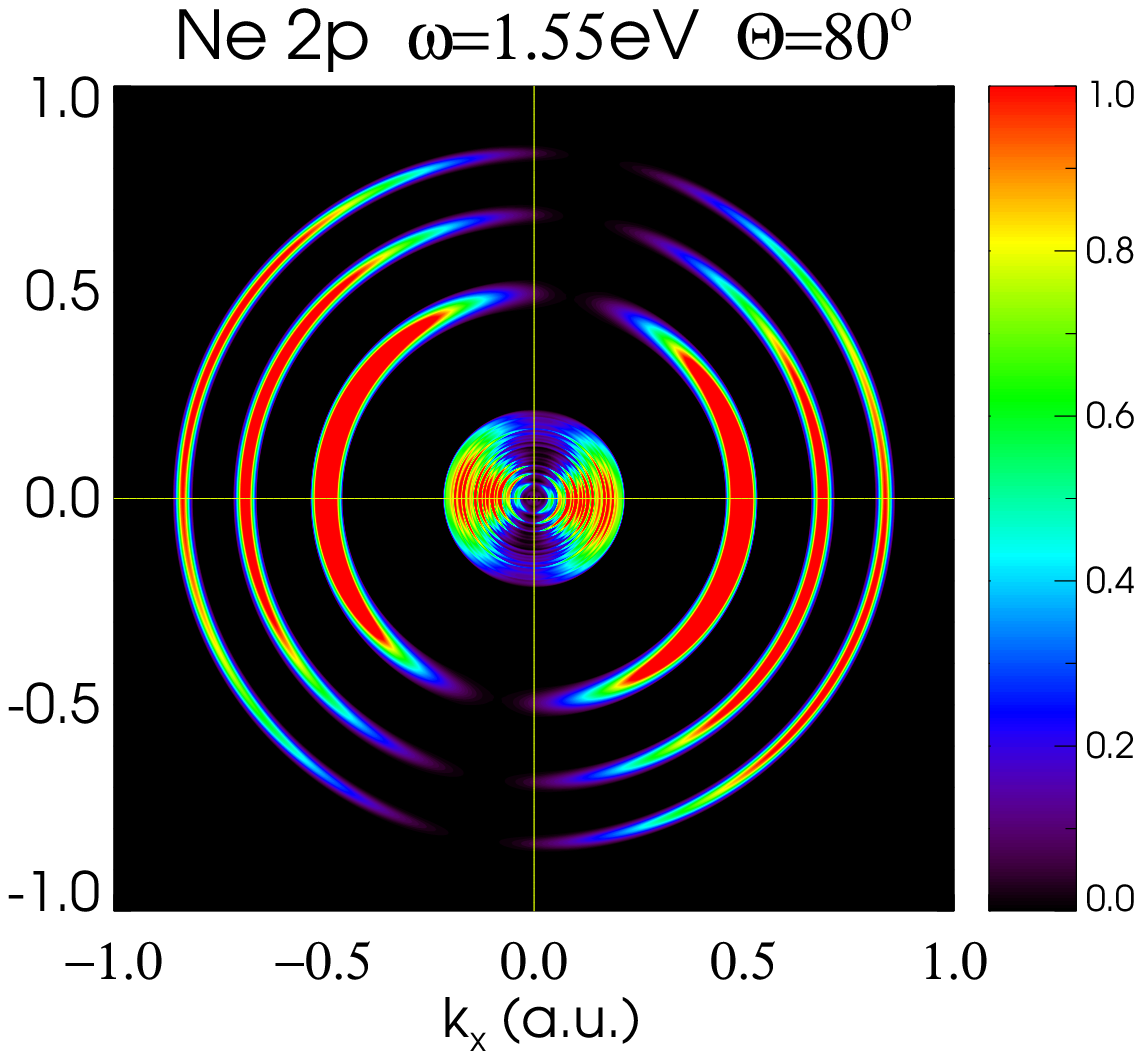}

\hs{-0.3cm}
\epsfxsize=16.5cm
\epsffile{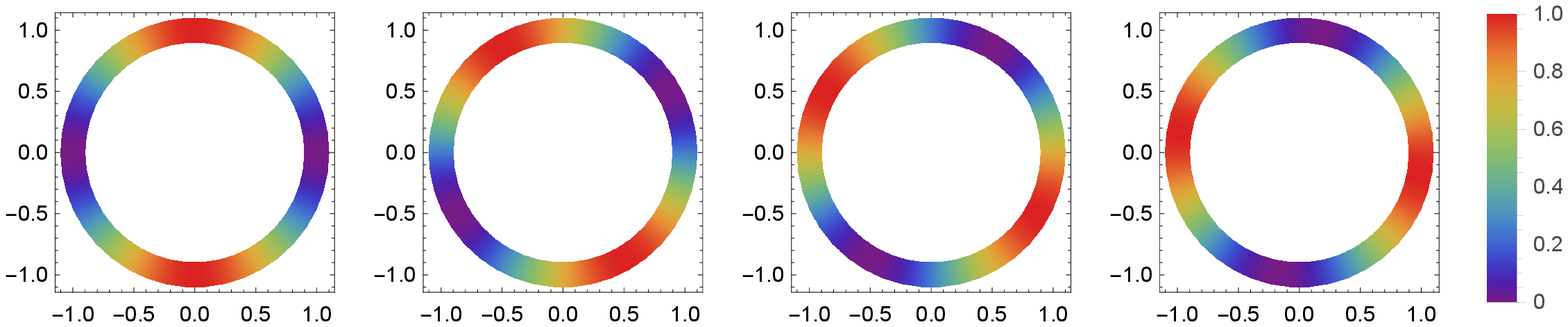}

\ep

\caption{Top row: PMD of neon projected on the joint XUV/IR
  polarization plane at various angles $\Q$. The XUV/IR time delay
  $\t=0$ in all cases. The SB orders are as
  marked. Bottom row: graphical visualization of the angular factor
  $\cos^2(\q-\Q)$.
\label{Fig4}}
\end{figure}

\np
\subsection{Neon}

Polarization control of RABBITT in Ne is demonstrated in
\Fref{Fig4}. Here we display the filtered PMD \eref{filter} projected
on the joint XUV/IR polarization plane at various angles
$\Q$. Similarly to \Fref{Fig2}, the prime harmonic peaks are masked
for clarity. These peaks show almost perfectly circular structure as
the angular anisotropy is very small in Ne $|\b|\ll1$ as is seen from
the left panel of \Fref{Fig1}. The SB's follow faithfully the
$\cos^2(\q-\Q)$ pattern as prescribed by the SPA.

\begin{figure}[h]
\bp{20cm}
\hs{-1cm}
\epsfxsize=6cm
\epsffile{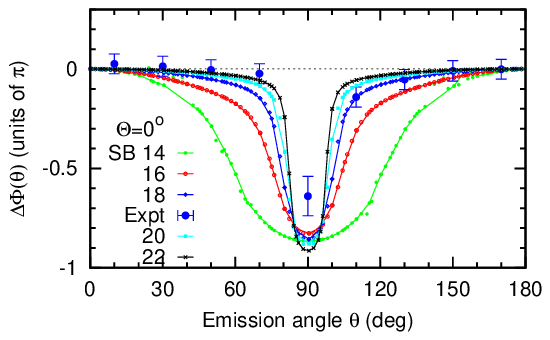}
\hs{-1cm}
\epsfxsize=6cm
\epsffile{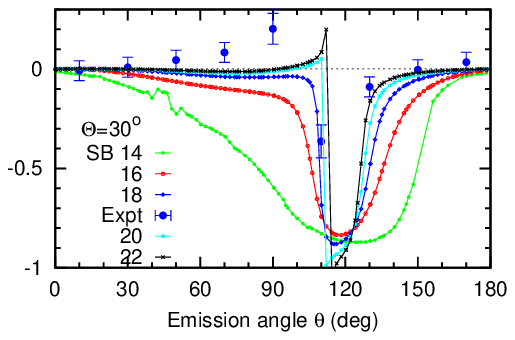}
\hs{-1cm}
\epsfxsize=6cm
\epsffile{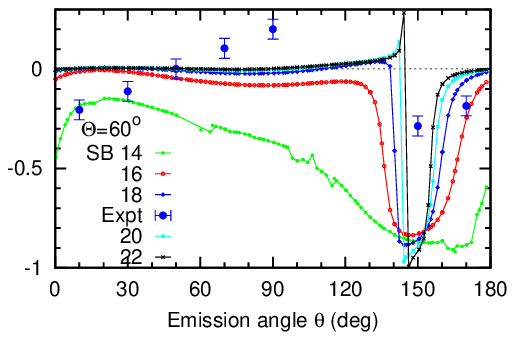}

\hs{-1cm}
\epsfxsize=6cm
\epsffile{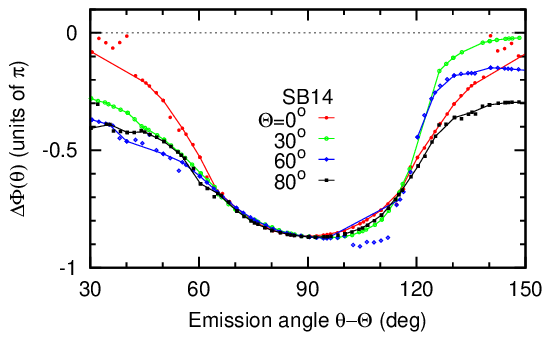}
\hs{-1cm}
\epsfxsize=6cm
\epsffile{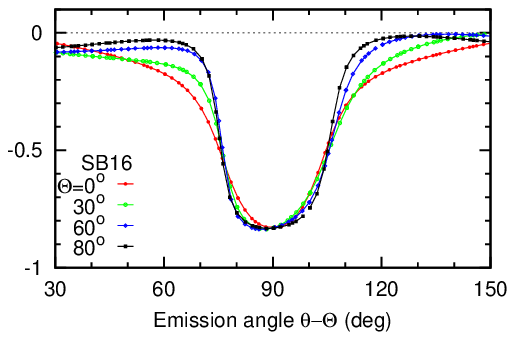}
\hs{-1cm}
\epsfxsize=6cm
\epsffile{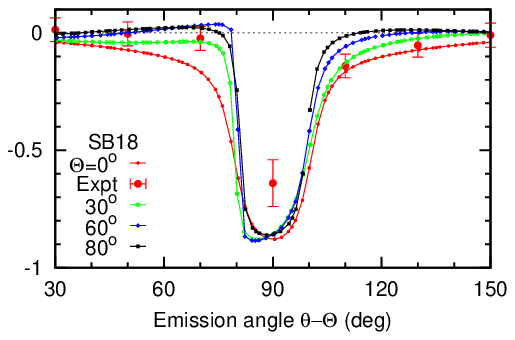}

\hs{-1cm}
\epsfxsize=6cm
\epsffile{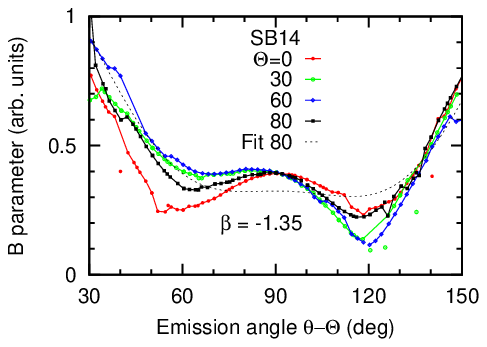}
\hs{-1cm}
\epsfxsize=6cm
\epsffile{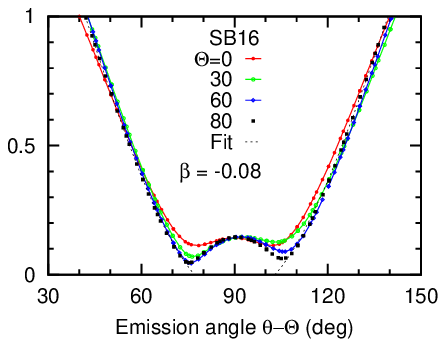}
\hs{-1cm}
\epsfxsize=6cm
\epsffile{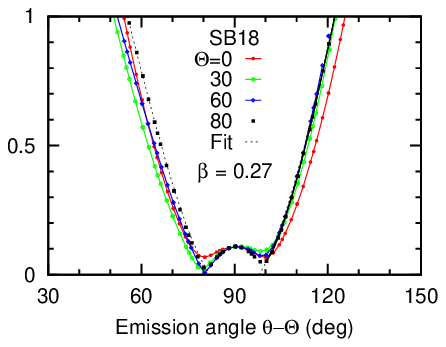}

\ep

\caption{Same as \Fref{Fig3} for neon. Top row: $\Delta\Phi$ is shown
  for various SB orders at a fixed polarization control angle
  $\Q$. The experimental data of \cn{Jiang2022} for SB18 at the
  polarization control angles $\Q=0$, $20^\circ$ and $54.7^\circ$ are
  shown with error bars.  Middle row: $\Delta\Phi$ is shown for
  several fixed SB orders while $\Q$ angle varies. The horizontal
  emission angle scale is shifted as $\q-\Q$.  Bottom row: $B$
  parameter is shown for several fixed SB orders while $\Q$ angle
  varies. The dotted line shows the analytic fit with $[1+\b
    P_2(\cos\q)]\cos^2(\q-\Q)$ for $\Q=80^\circ$. The corresponding
  $\b$ values are displayed.
\label{Fig5}}
\end{figure}

The polalrization control of the angular dependent RABBITT phase and
magnitude parameters in Ne is demonstrated in \Fref{Fig5}. As in
\Fref{Fig3}, the top row of panels shows $\Delta\Phi(\q)$ for a range
of SB orders at various fixed $\Q$ values. In the middle row of
panels, we group $\Delta\Phi(\q)$ for a given SB order while the
polarization control angle $\Q$ varies. However, unlike in the middle
panel of \Fref{Fig3}, we shift the horizontal axis by the whole angle
$\Q$ rather than by $\Q/2$. Thus achieved phase centering near
$\q-\Q=90^\circ$ is nearly perfect.
Such an  angular centering of the RABBITT phase in Ne can
be understood within SPA.  We note from the left panel of \Fref{Fig1}
that $|\b|\ll1$ for SB16 and SB18. Hence the angular dependence of the
RABBITT spectrum is given by the IR factor $\cos^2(\q-\Q)$ while the
XUV angular factor remains flat.

We observe that SB14 displays a significantly broader angular
variation which starts much further away from the center
$\q-\Q=90^\circ$. This is a manifestation of the uRABBITT process in
Ne which is responsible for the formation of SB14 as the PP13 falls
below the threshold. As in the case of He, the experimental data by
\cn{Jiang2022} for SB18 are close to our calculation in the collinear
case but deviate significantly as the polarization control angle $\Q$
increases. We also note that the experimental data for Ne do not
adhere to the $\q-\Q=90^\circ$ centering whereas the accompanying
$R$-matrix calculation does display this centering.

The bottom row of panels in \Fref{Fig5} displays the angular dependent
$B$ paramters on the adjusted angular scale. On this scale, the $B$
parameters are centered relative to the angle $\q-\Q=90^\circ$. This
centering is more accurate in SB16 and 18 in comparison with the
uRABBITT SB14. The latter is only centered in a close vicinity of
$90^\circ$.  In the SB16 formed by the uRABBITT process, the $B$
parameter never falls close to zero. Accordingly, the phase varies
rather smoothly on a wide angular range. In the conventional RABBITT SB16 and SB18, the magnitude $B$
falls steeply towards the center $\q-\Q=90^\circ$ and almost reaches
zero, especially for SB18. The $\b$ values extracted from the analytic
fit with \Eref{soft} at $\Q=80^\circ$ are shown on each panel. For the
uRABBITT SB14 this fit produces an unphysical result $\b<-1$. This can
be understood as the $\b$ parameterization is not applicable to a
discrete-discrete XUV photon absorption responsible for the uRABBITT
process. For conventional SB16 and 18, the $\b$ values are reasonable
and close to those exhibited in the left panel of \Fref{Fig1}. As $\b$
grows, the angular variation of the $B$ and $C$ parameters become
sharper and centers closer to $\q-\Q=90^\circ$.

\np

\begin{figure}[ht]
\bp{20cm}
\hs{-2cm}
\epsfxsize=7cm
\epsffile{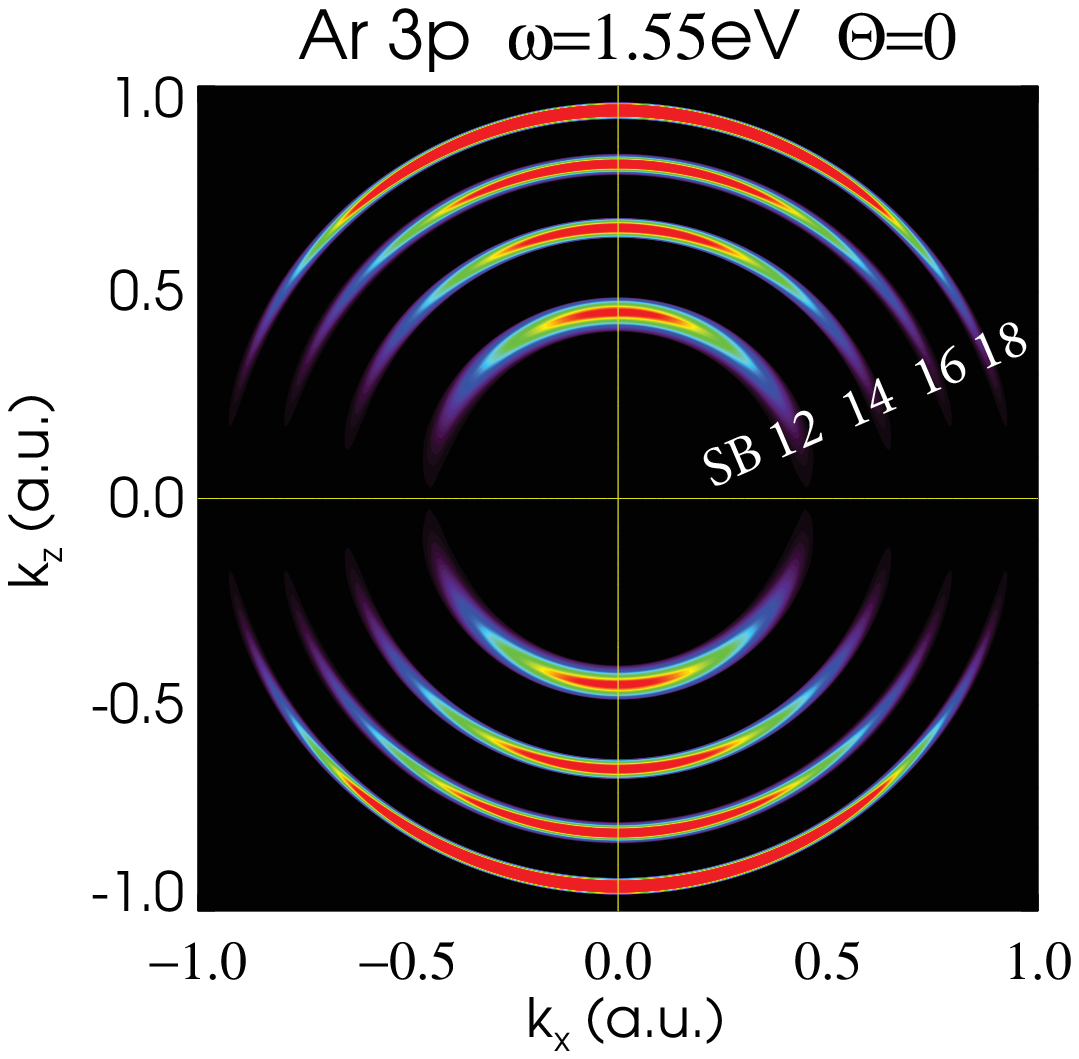}
\hs{-3.3cm}
\epsfxsize=7cm
\epsffile{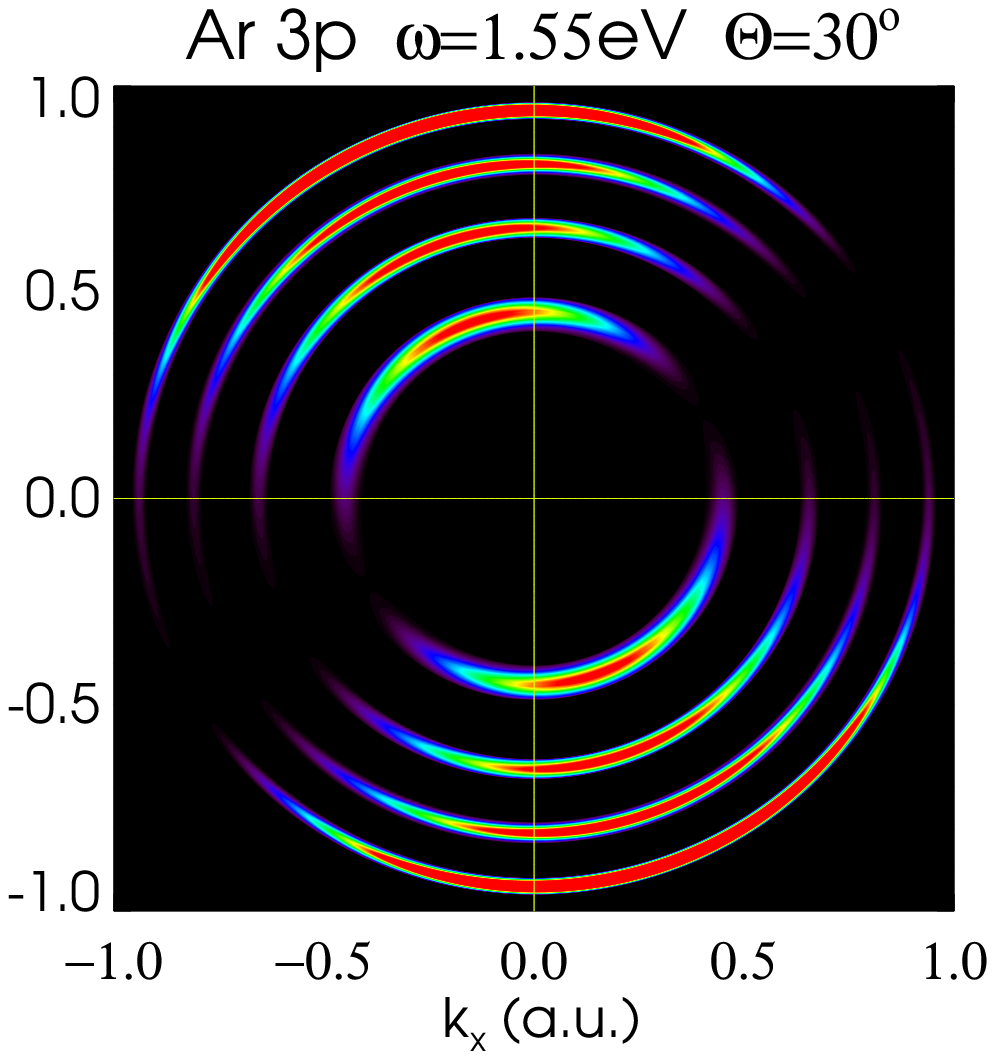}
\hs{-3.3cm}
\epsfxsize=7cm
\epsffile{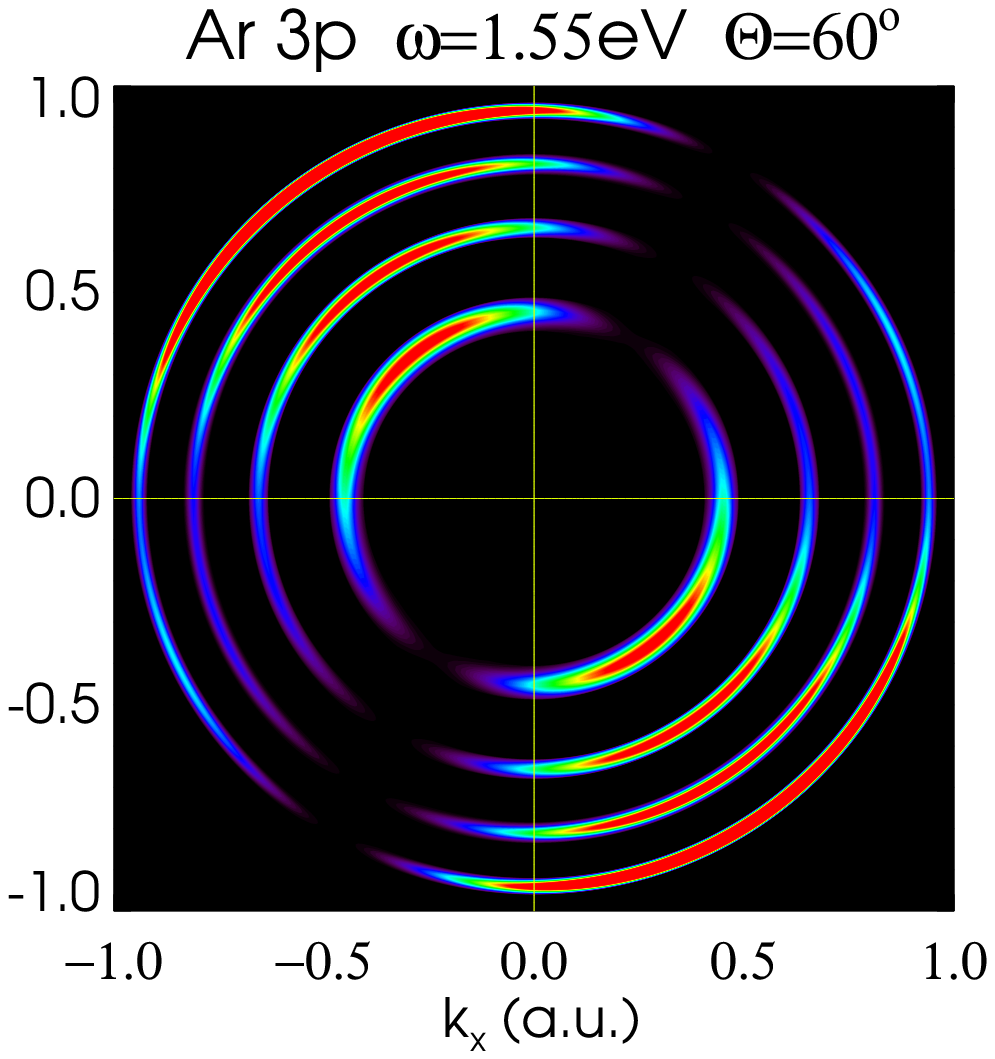}
\hs{-3.3cm}
\epsfxsize=7cm
\epsffile{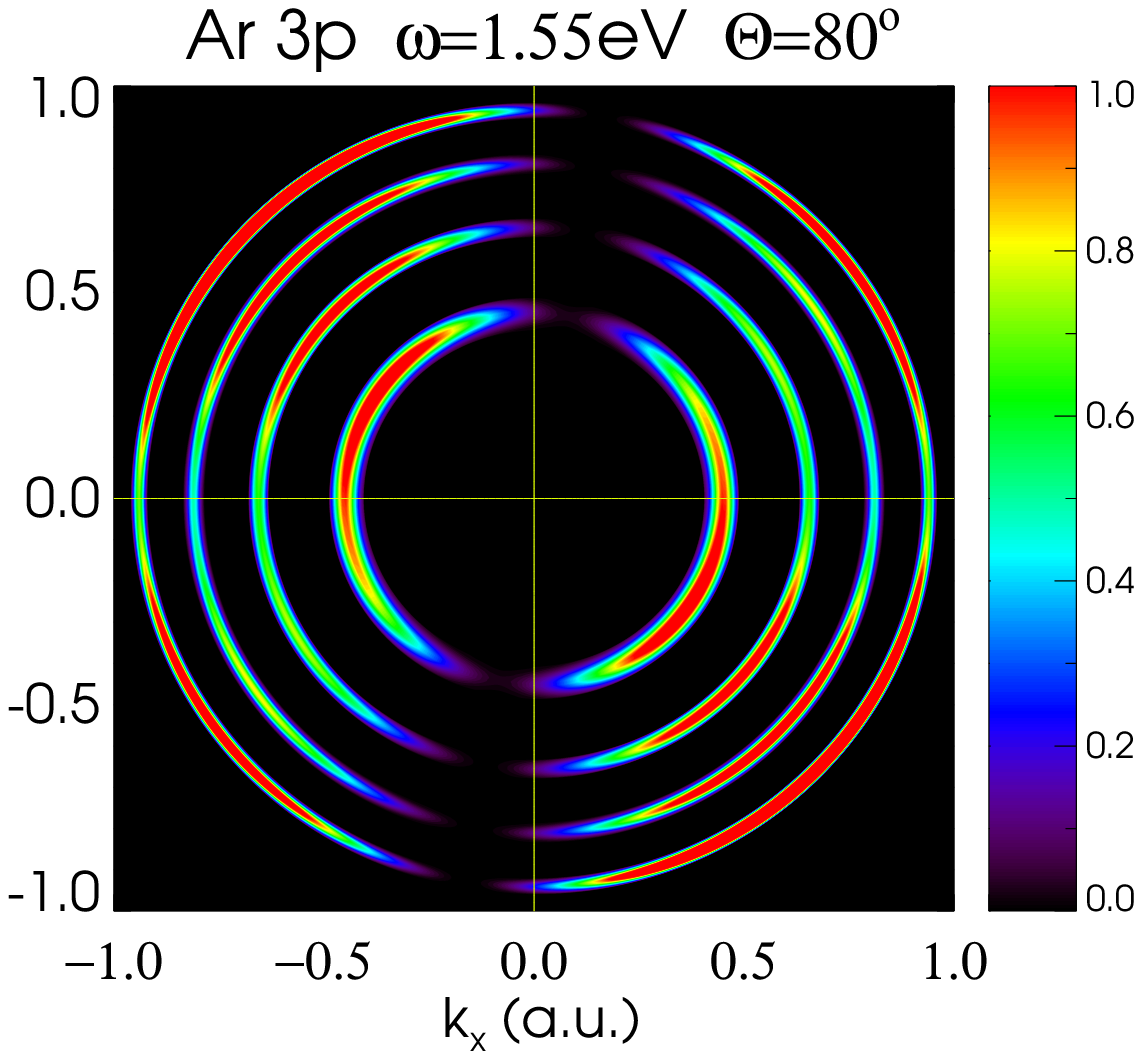}

\hs{-0.3cm}
\epsfxsize=16.5cm
\epsffile{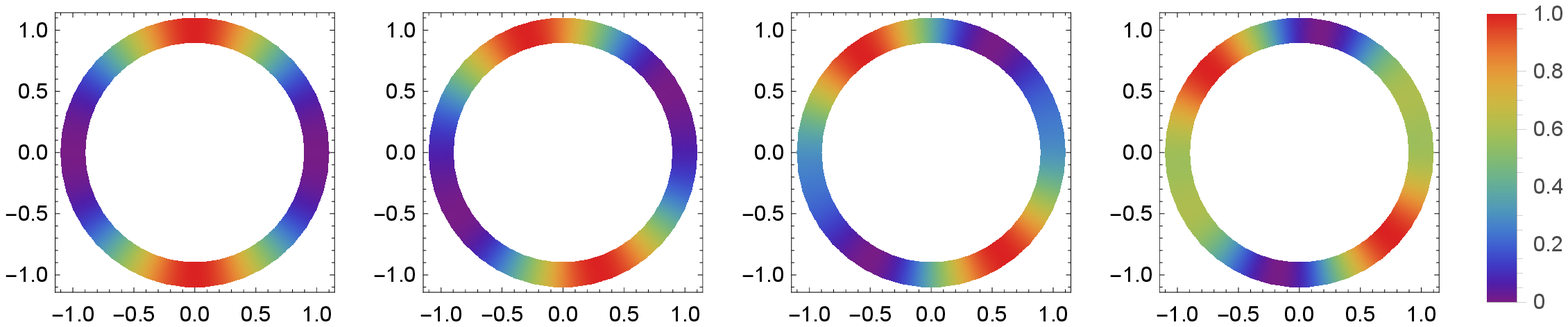}

\ep

\caption{Top row: PMD of argon projected on the joint XUV/IR
  polarization plane at various angles $\Q$. The XUV/IR time delay
  $\t=0$ in all cases. The SB orders are as marked. Bottom row:
  graphical visualization of the angular factor $[1+\beta
    P_2(\cos\q)]\cos^2(\q-\Q)$ with $\b=1$.
\label{Fig6}}
\end{figure}

\subsection{Argon}

\begin{figure}[h]
\bp{20cm}
\hs{-1cm}
\epsfxsize=6cm
\epsffile{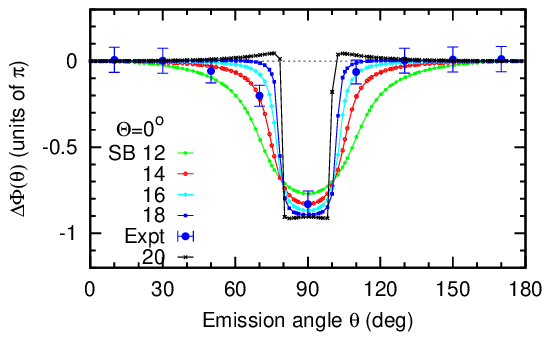}
\hs{-1cm}
\epsfxsize=6cm
\epsffile{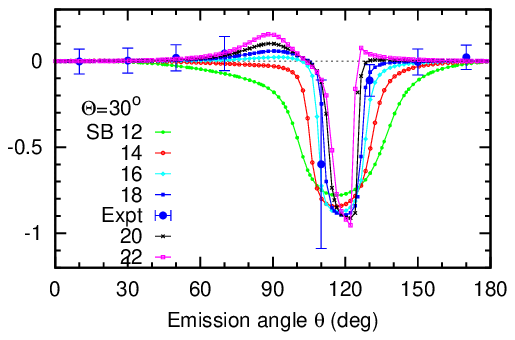}
\hs{-1cm}
\epsfxsize=6cm
\epsffile{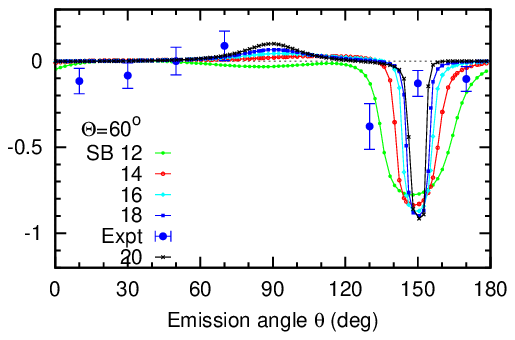}

\hs{-1cm}
\epsfxsize=6cm
\epsffile{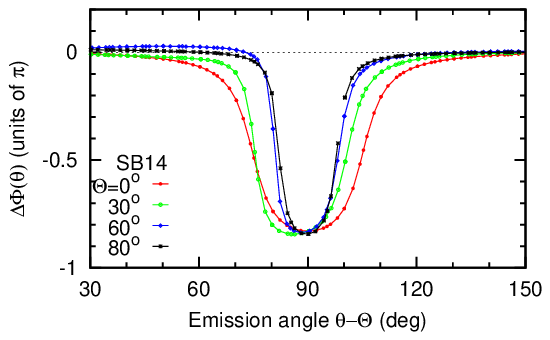}
\hs{-1cm}
\epsfxsize=6cm
\epsffile{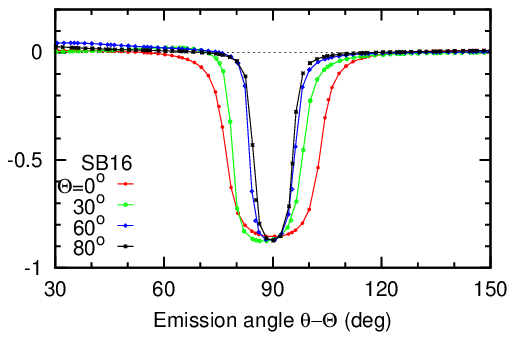}
\hs{-1cm}
\epsfxsize=6cm
\epsffile{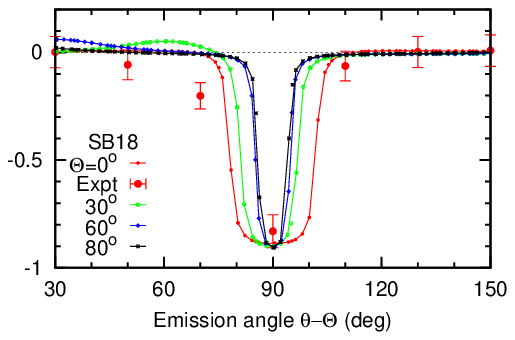}

\hs{-1cm}
\epsfxsize=6cm
\epsffile{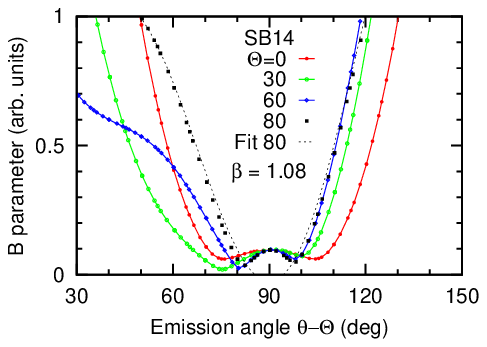}
\hs{-1cm}
\epsfxsize=6cm
\epsffile{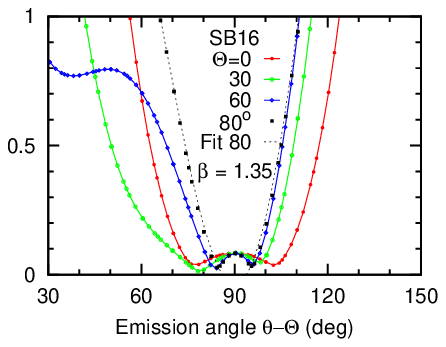}
\hs{-1cm}
\epsfxsize=6cm
\epsffile{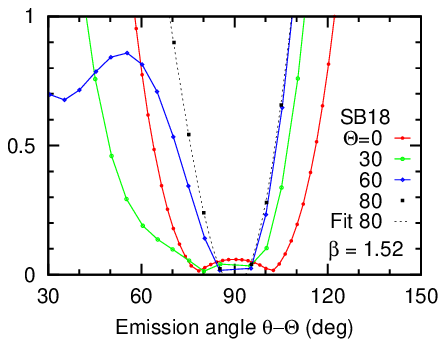}

\ep

\caption{ Same as \Fref{Fig5} for argon. Top row: $\Delta\Phi$ is
  shown for various SB orders at a fixed polarization control angle
  $\Q$. The experimental data of \cn{Jiang2022} for SB18 at the
  polarization control angles $\Q=0$, $20^\circ$ and $54.7^\circ$ are
  shown with error bars. Middle row: $\Delta\Phi$ is shown for several
  fixed SB orders while $\Q$ angle varies. The horizontal emission
  angle scale is displaced as $\q-\Q$.  Bottom row: $B$ parameter is
  shown for several fixed SB orders while $\Q$ angle varies. The
  dotted line shows the analytic fit with $[1+\b
    P_2(\cos\q)]\cos^2(\q-\Q)$ for $\Q=80^\circ$. The corresponding
  $\b$ values are displayed.
\label{Fig7}}
\end{figure}

The polarization control of RABBITT in Ar is demonstrated in
Figs.~\ref{Fig6} and \ref{Fig7}. In the case of Ar, all the considered
sidebands are formed by a conventional RABBITT process. The
under-threshold uRABBITT in Ar can be observed at 400~nm
\cite{Kheifets2023} but does not manifest itself in the present case
at 800~nm . As is seen from the right panel of \Fref{Fig1},
$\b\simeq1$ for the shown SB's in Ar. Accordingly, the PMD follows the
$[1+ P_2(\cos\q)]\cos^2(\q-\Q)$ angular pattern which is displayed on
the bottom panel of \Fref{Fig6}.

The angular dependent RABBITT parameters in Ar are visualized in
\Fref{Fig7}. The top row of panels displays the phases of several SB's
at a fixed polarization control angle $\Q$. In the middle panel, the
SB order is fixed while the angle $\Q$ varies. Similarly to Ne, the
RABBITT phases are perfectly centered relative to the angle
$\q-\Q=90^\circ$. The width of the angular variation relative to this
angle  decreases systematically as the angle $\Q$ grows. The
experimental data by \cn{Jiang2022} agree well with our calculation in
the collinear case but they become too sparse to reproduce a rather
narrow angular variation near $\q-\Q=90^\circ$ for larger $\Q$. In the
meantime, the theoretical results accompanying the experiment by  
\cn{Jiang2022} conform to the $\q-\Q=90^\circ$ centering. 

\np
The bottom row of panels of \Fref{Fig7} shows the angular variation of the
magnitude $B$ parameters which are fitted with \Eref{soft} at
$\Q=80^\circ$. The corresponding $\b$ parameters agree well with other
theoretical and experimental values displayed in the right panel of
\Fref{Fig1}.  As $\b$ increases, the angular dependence of the $B$ and
$C$ parameters sharpens, especially at larger polarization control
angles $\Q$.

\section{Conclusion}
\label{Conclusion}

In the present work we study systematically the polarization control
of RABBITT in noble gas atoms, from He to Ar. As the control variable,
we use the mutual angle $\Q$ formed by the non-collinear XUV/IR
polarization axes.  Our analysis is based on the numerical solution of
the TDSE driven by a combination of XUV and IR pulses at a variable
delay. We also invoke the LOPT and SPA to interpret  our
numerical results qualitatively.

We visualize the polarization control of RABBITT by two sets of
graphical presentations. In the panoramic view, we display the PMD
projected on the joint XUV/IR polarization plane and highlight the SB
angular symmetry by masking the normally stronger prime harmonic
peaks. While the latter peaks are insensitive to the polarization
control, the SB's can be easily manipulated. Their angular symmetry
follows closely the predictions of SPA for all target atoms. 

In a more detailed view, we analyze the angular dependent magnitude
and phase of the RABBITT oscillations. This analysis is greatly
simplified by an angular shift of the photoelectron emission angle
$\q$ which centers the RABBITT parameters relative to the angle
$\q-\Q/2=90^\circ$ in H and He and that of $\q-\Q=90^\circ$ in Ne and
Ar. Such a centering is explained by the corresponding values of the
angular anisotropy $\b$ parameter. While $\b=2$ in $s$-electron
targets, $\b\ll1$ in Ne and $\b\lesssim1$ in Ar at small photoelectron
energies that we analyze.  The experimental data of \cn{Jiang2022} are
too sparse to subject them to a rigid centering test whereas the
accompanying theory as well as the calculations by \cn{Boll2023} do
comply with this test. We hope that experimental data an a denser
angular grid with better statistics will appear soon to conduct a more
conclusive test. We also plan to extend our studies towards the
diatomic molecules such as \H in which the PMD demonstrates much more
complicated angular structure \cite{Serov2017,Serov2023}. We expect
that the polarization control of RABBITT in \H would provide a deeper
insight into this process.

\section*{Acknowledgment} 

We gratefully acknowledge support of the National Computational
Infrastructure facility (NCI Australia) which was instrumental for
this work.

\np
\section*{Bibliography} 


\end{document}